\documentstyle{mn}
\begin{document}

\title{Non-pulsed gamma radiation from binary system with a pulsar}
\author[M.A.Chernyakova, A.F.Illarionov]{M.A.Chernyakova\thanks{
masha@sigma.iki.rssi.ru}, A.F.Illarionov \\
Astro Space Center, P. N. Lebedev Physical Institute, 84/32 Profsoyuznaya
St., Moscow 117810, Russia}
\date{}
\maketitle

\begin{abstract}
Consider binary system with a millisecond pulsar ejecting relativistic
particles and an optical star emitting soft photons with the energy $\omega
\simeq 1-10$ eV. These low-energy photons are scattered by the relativistic
electrons and positrons of the pulsar's wind. The scattered photons forms a
wide spectrum from hard X-ray band up to gamma band $\varepsilon \simeq
1-1000$GeV .When the pulsar wind is isotropic the luminosity of gamma
radiation $L_\gamma =L_\gamma (\psi )$ depends heavily on the angle $\psi $
between the directions to the optical star and to the observer from the
pulsar. During the orbital motion this angle varies periodically giving rise
to the periodical change of the observed intensity of the gamma radiation
and its spectrum. We calculated the spectral shape of the scattered hard
photons. Under the assumption that the energy losses of the relativistic
particles are small we receive analytical formulas. We apply our results to
the binary system PSR\ B1259-63 and show that if the wind from the Be star
is accounted for then it is possible to reproduce the observed spectrum.
\end{abstract}


\section{ Introduction}

The principal accessible store of energy in a pulsar is its rotational
energy, which is liberated at a rate $L_p\simeq I\stackrel{.}{p}4\pi
^2/p^3\sim 10^{32}\div 10^{36}$erg/s ( see, for example, a table of pulsars
parameters ( $L_p,p,$ $\stackrel{.}{p}$) given in the book of Beskin et
al.,1993 ), where $I$ $\simeq 10^{45}$g cm$^{-2}$ is moment of inertia, $p$
is a period, $\stackrel{.}{p}$ is a deceleration of a pulsar. The bulk of
the pulsar energy $L_p$ is transferred to the pulsar wind which consists of
electrons, positrons and probably heavy ions, $L_w=gL_p$, $g\leq 1.$ The
Lorentz factor of the relativistic particles in the wind may vary in range $%
\gamma \sim 10-10^6$ (e.g. Manchester\&Taylor,1977). The intrinsic gamma ray
luminosity of pulsed emission from the short periodic pulsars is of the
order of $L_\gamma /L_p\approx 0.01$ (see Arons, 1991). Below we discuss the
different (induced) mechanism of non pulsed gamma radiation generation in
binary with a pulsar and an optical star. This mechanism could result in
considerably higher ratio of $L_\gamma /L_p$. The preliminary results of
this work were published in the paper of Chernyakova\&Illarionov, 1997.

Consider the case of binary with a pulsar ejecting relativistic particles
and an optical star emitting soft photons in optic and UV band with the
energy $\omega \simeq 1-10$ eV. These low-energy photons are scattered by
the pulsar wind relativistic electrons and positrons. The energy of the
photon after the inverse Compton scattering is very high - $\varepsilon
_{\max }\sim \omega \gamma ^2$ in the Thomson limit $(\omega \gamma \ll
mc^2) $ and $\varepsilon _{\max }\sim mc^2\gamma $ in the opposite case ( $m$
is a mass of an electron, $c$ is a light velocity). The scattered photons
form a wide spectrum from hard X-ray band up to gamma band $\varepsilon
\simeq 1$ $-1000$GeV. The relativistic particle scatters the photon
preferably along the direction of the particle velocity. As a result while
soft photons are directed radially from the optical star, the scattered hard
photons move radially from the pulsar. Here and bellow we assume that the
pulsar wind particles are radially directed from the pulsar inside the
effective scattering volume.

In the case of the presence of an obstacle for the pulsar wind and mainly
when the matter flow from the optical star is rather intensive the radial
flow of the relativistic wind is destroyed. Then the system of shock waves
resulting from the collision of two winds appears between the pulsar and the
optical star and the trajectories of the electrons and positrons beyond the
shock change. In the work of Tavani and Brookshaw (1991) the case of weak
(in comparison with the pulsar wind) matter outflow is discussed. The
hydrodynamics of the collision between the relativistic and the
nonrelativistic winds is closely analogous to the hydrodynamics of the two
nonrelativistic winds collision, which was intensively discussed by
different authors applied to the WR+OB binaries since the work of
Prilutskii\&Usov (1976).

The total luminosity of scattered hard photons $L_\gamma $ is equal to the
total particle energy losses $L_{loss}$ through the inverse Compton
scattering in the course of the particle motion from the pulsar to the
infinity. To estimate $L_\gamma $ we note that the rate of the energy losses
of a relativistic particle moving in a radiation field with energy density
of soft photons $w_{soft}\simeq L_{*}/4\pi a^2c$ is about $mc^2d\gamma
/dt\simeq -w_{soft}\sigma _Tc\gamma ^2$ in the Thomson limit. Here $L_{*}$
is a luminosity of the optical star, $a$ is a binary stars separation and $%
\sigma _T=\frac{8\pi }3\left( e^2/mc^2\right) ^2=6.65\times 10^{-25}$cm$^2$
is the Thomson cross-section. Hence the decrease of the Lorentz factor of a
particle is $\Delta \gamma \sim \gamma \left[ 1-\left( 1+\gamma /\gamma
_{*}\right) ^{-1}\right] ,$ $\gamma _{*}=4\pi amc^3\left/ \sigma
_TL_{*}\right. $. So the rate of all particles total energy loss is

\[
L_\gamma =L_{loss}=\frac{L_w\Delta \gamma }\gamma \sim L_w\left( 1-\frac
1{1+\gamma /\gamma _{*}}\right) .
\]
Thus in the case $\gamma \ll \gamma _{*}$ the gamma-ray luminosity from
binary is proportional to the luminosity of the optical star $L_{*}$ and to
the luminosity of the wind $L_w$. $L_\gamma =KL_w$, where the transformation
parameter $K=\gamma /\gamma _{*}$. In the close binary system with an
optical star with high luminosity $\gamma \gg \gamma _{*}.$ In this case
practically all energy of the wind transfers to the energy of scattered
photons $L_{\gamma \max }\approx L_w$.

In each unit of the volume in the region where the gamma radiation is
generated the source function of gamma radiation is highly anisotropic. When
the pulsar wind is isotropic the gamma luminosity $L_\gamma (\psi )$ ( $%
L_\gamma =\int 2\pi L_\gamma (\psi )d\cos \psi $) has an azimuthal symmetry
around the line connected binary companions and depends heavily on the angle
$\psi $ between the directions to the optical star and to the observer from
the pulsar, see Figure 1. In our case of free radial relativistic wind the
binary system emits the maximum energy of gamma radiation in the direction
of the star $(\psi =0)$ and the minimum energy of radiation in the opposite
direction ($\psi =\pi )$. The spectrum of the radiation and the maximal
radiated energy also depend on $\psi $. During orbital motion $\psi $ varies
periodically giving rise to the periodical change of the intensity of the
gamma radiation coming from the binary system to the observer.

In section 2 we calculate the spectral shape $L_\gamma (\varepsilon ,\psi )$
of the scattered hard photons in the case of arbitrary value of the
parameter $\frac{\omega \gamma }{mc^2}$, going beyond the Thomson limit.
Under the assumption $K\ll 1$ we receive analytical formula for the $%
L_\gamma (\varepsilon ,\psi )$.

In section 3 we apply our results to the binary system PSR\ B1259-63 and
find that under the assumption of the power law relativistic particles
spectrum, $\frac{dN_{e^{\pm }}}{d\gamma }=0.4L_w\left/ \left[ mc^2\left(
\gamma _{\min }^{-0.4}-\gamma _{\max }^{-0.4}\right) \right] \right. \gamma
^{-2.4}$ in the range $10<\gamma <500$, our model describes the observe
photon spectrum rather good but the intensity is less then the observed one
by a factor about 30. This discrepancy is due to the presence of the mass
outflow from the Be star which disturbed the free flow of the pulsar wind.
The centrally located shock appears between the pulsar and the star due to
the interaction between the two winds. The big differences between the
values of the velocities of the particles from the different sides of the
tangential discontinuity will lead to the growth of the instabilities and
the two winds will be macroscopically mixed between the shocks. Then the
heavy non relativistic wind slows down the volumes filled by the
relativistic electrons and positrons and they acquire essentially non
relativistic hydrodynamic drift velocity $v_d$ along the shock while the
energy of electrons and positrons does not changes significantly. With the
decrease of the hydrodynamic velocity of the relativistic plasma the time
which it spends near the optical star increases in $c/v_d$ times. The
effective transformation parameter $K_{eff}\sim \frac c{v_d}K$ thus can be
large enough to overcome the discrepancy between the simple theory and
observations.

\section{Generation of gamma radiation}

Lets us consider an interaction between the relativistic pulsar wind and the
soft radiation from the companion and find the spectral and angular
dependence of the outgoing hard radiation $L_\gamma (\varepsilon ,\psi )$.
We assume that the pulsar wind and the soft emission from the companion are
isotropic and that there is no mass outflow from the optical star. We treat
both the pulsar and the optical star as a point sources. In this case the
trajectories of the particles and soft non scattered photons are directed
radially from the pulsar and from the optical star correspondingly.

Let I denote the location of the small element of volume $dV$ at a distance $%
r$ to the pulsar and at an angle $\theta _2$ to the line of sight $%
\overrightarrow{PO}$ (see Figure 1). In this element of volume optical
photons with the energy $\omega $ scatters by the relativistic particles
(electrons or positrons). The angle between the directions of the photon and
the particle movements before the interaction is denoted by $\theta _1$. The
photon scattering angle is designated by $\theta $. Lets $l$ signifies the
distance from this volume to the companion $S$.

We are interested in the case of a distant observer and thus a vector along
the direction to the observer from any point near the system may be
considered as parallel to such a vector from the pulsar. Only the photons
scattered in the direction of the observer within the small solid angle $%
\Omega =S\cos \chi /D^2$ will reach the observer. Here $S$ is the area of
the observer's surface, $D$ is a distance from the pulsar to the observer
and $\chi $ is an angle between the direction to the observer from the
pulsar and from the volume. As $\chi $ is of the order of $a/D\ll 1$, we may
neglect the variance of $\Omega $ from point to point. As we have already
mention relativistic particle scatters the photons preferably along the
direction of the particle velocity and thus $\theta _2\ll 1$. From this fact
it follows that angles $\theta $ and $\theta _1$ depend only on $r$ and with
an accuracy of the order $\theta _2\sim \sqrt{\omega /\varepsilon }$ are
equal.

\subsection{Anisotropy of the energy radiated from the binary system in the
Thomson limit.}

Lets calculate the $\psi $-dependence of the luminosity of the hard
radiation emitted from the binary system $L_\gamma (\psi )$ in the Thomson
limit ( $\omega \gamma /mc^2\ll 1$ ). The luminosity of scattered hard
photons $L_\gamma (\psi )$ is equal to the particle energy losses in the
course of the particle motion from the pulsar to the infinity. So at first
lets calculate the total energy losses $L_{lossT}$ of the electron moving
along the radial trajectory at an angle $\psi $ to the line $PS$ in the
radiation field of the optical star. The rate of the relativistic ( $\gamma
\gg 1$ ) electron's energy losses is given by

\begin{equation}
\frac 1c\frac{d\gamma }{dt}=\frac{d\gamma }{dr}=-\frac{L_{*}\sigma _T\gamma
^2}{4\pi mc^3}\frac{\left( 1-\cos \theta _1\right) }{l^2}^2,  \label{av}
\end{equation}
where 
\[
l=\sqrt{\left( r-a\cos \psi \right) ^2+a^2\sin ^2\psi ,} 
\]
\[
\cos \theta _1=\left. \left( r-a\cos \psi \right) \right/ l. 
\]

Upon integrating (\ref{av}) over radius from $r=0$ to $r$ we obtain the
dependence of the Lorentz factor of the electron on $r$ :

\begin{eqnarray}
\gamma &=&\gamma _0\left[ 1+K\Phi (\psi )+\right.  \nonumber \\
&&\left. \frac K{\sin \psi }\left( \frac 2{\sqrt{1+x^2}}-\frac 32arccot{x}%
-\frac x{2\left( 1+x^2\right) }\right) \right] ^{-1},  \label{gam}
\end{eqnarray}
where

\begin{equation}
\Phi (\psi )=\frac{3(\pi -\psi )}{2\sin \psi }-2-\frac{\cos \psi }2
\end{equation}
is a beaming function which as it will follows from ( \ref{de} ) represents
the anisotropy of the gamma radiation, 
\begin{equation}
K=\gamma _0/\gamma _{*}=\sigma _TL_{*}\gamma _0\left/ (4\pi amc^3)\right.
\end{equation}
represents the efficiency of the electron energy transformation to the
energy of gamma radiation, the parameter $x$ is determined as 
\[
x=\frac{r/a-\cos \psi }{\sin \psi } 
\]
and $\gamma _0$ is the initial Lorentz factor of the electron. Letting $%
r\rightarrow \infty $ we find from (\ref{gam}) the total change of the
Lorentz factor $\Delta \gamma (\psi )=\gamma _0-\gamma (x=\infty ,\psi
)=\gamma _0\left[ 1-\left( 1+K\Phi (\psi )\right) ^{-1}\right] $.

As the bulk of the energy losses of the relativistic electron transfers to
the scattered photons moving in the direction of the electron movement the
angular dependence of the energy radiated from the binary system $L_\gamma
(\psi )$ agree with the angular dependence of the energy losses of the
electrons. The total energy losses of the isotropic flow of all relativistic
particles in the direction of the observer located at an angle $\psi $ to
the line $PS$ in a small solid angle $\Omega $ are proportional to the
amount of particles moving in this direction $N_{e_{\pm }}\Omega /4\pi
=L_w\Omega /4\pi \gamma _0mc^2$ in the case of monoenergetic pulsar wind,
and to the total energy losses of one particle $mc^2\Delta \gamma (\psi )$ 
\begin{eqnarray}
L_\gamma (\psi )\Omega &\approx &L_{lossT}(\psi )\Omega =\frac{L_w\Delta
\gamma (\psi )}{4\pi \gamma _0}\Omega =  \nonumber \\
&&\ \ \ \ \frac{L_w}{4\pi }\left( 1-\frac 1{1+K\Phi (\psi )}\right) \Omega .
\label{de}
\end{eqnarray}
This is the energy transferred to the gamma band. Figure 2 shows the
dependence $\Phi (\psi )$. The beaming function $\Phi (\psi )$ is rapidly
decreasing function. 
At $\psi \ll 1$ $\Phi (\psi )\simeq \pi /\psi $ and at $\pi -\psi \ll 1$ $%
\Phi (\psi )\simeq (\pi -\psi )^4/120$. Equation (\ref{de}) is valid for the
arbitrary value of $K$ but it has no use for $\psi $ less then $R_{*}/a$ and
bigger then $\pi -R_{*}/a$ ($R_{*}$ - is a radius of the optical star) as
the optical star is not a point source. That's why we don't worry about the
fact that $\Phi (\psi )\rightarrow \infty $ as $\psi \rightarrow 0$. If $%
\psi $ is not too small then it follows from (\ref{de}) that under the
condition $K\ll 1$ the energy decrease is small and 
\begin{equation}
L_\gamma (\psi )=\frac{L_w\gamma _0}{4\pi \gamma _{*}}\Phi (\psi ).
\label{de1}
\end{equation}
Upon integrating (\ref{de1}) over $\psi $ from $\psi =0$ to $\psi =\pi $ we
obtain the total energy transferred to the gamma band in the case $K\ll 1$

\begin{equation}
L_\gamma =\left( \frac 38\pi ^2-2\right) \frac{\gamma _0L_w}{\gamma _{*}}.
\label{ltot}
\end{equation}

\subsection{The spectral and the angular dependence of the outgoing radiation
}

Now lets calculate the spectral and the angular dependence of the outgoing
radiation $L_\gamma (\omega ,\varepsilon ,\gamma ,\psi )$ in the case of
arbitrary value of the parameter $\omega \gamma /mc^2$, going beyond the
Thomson limit. The luminosity of the scattered quanta moving at an angle $%
\psi $ to the line $PS$ within a unit solid angle in a unit of time in the
energy range from $\varepsilon $ to $\varepsilon +d\varepsilon $ may be
written in the following form: 
\begin{equation}
L_\gamma \left( \omega ,\varepsilon ,\gamma ,\psi \right) d\varepsilon =\int
\varepsilon n_{e^{\pm }}n_\omega c\left( 1-\beta \cos \theta _1\right)
\sigma _KdV,  \label{def}
\end{equation}
where $\beta =v/c$, $n_{e^{\pm }}$ is the density of the particles, $%
n_\omega =$ $L_{*}\left/ (4\pi cl^2\omega )\right. $ is the density of the
optical photons, $\sigma _K\left( \theta ,\theta _1,\theta _{2,}\omega
\right) $ - is the Klein-Nishina cross-section (Jauch \&Rohrlich 1976) 
\begin{eqnarray}
\sigma _K &=&\frac{3\sigma _T}{16\pi \gamma ^2}\frac{\varepsilon ^2}{\omega
^2}\left( 1-\beta \cos \theta _1\right) ^{-2}\times  \label{sigm} \\
&&\ \ \left[ \left( \frac{1-\cos \theta }{\gamma ^2\left( 1-\beta \cos
\theta _1\right) \left( 1-\beta \cos \theta _2\right) }-1\right) ^2\right. +1
\nonumber \\
&&\ \ +\left. \left( \frac \varepsilon {mc^2\gamma }\right) ^2\frac{\left(
1-\cos \theta \right) ^2}{\left( 1-\beta \cos \theta _2\right) \left(
1-\beta \cos \theta _1\right) }\right] .  \nonumber
\end{eqnarray}

The energy of the scattered photon is ( Berestetskii et al. 1971 )

\begin{equation}
\varepsilon =\omega \frac{1-\beta \cos \theta _1}{1-\beta \cos \theta
_2+\frac \omega {mc^2\gamma }\left( 1-\cos \theta \right) }.  \label{enscat}
\end{equation}
From (\ref{enscat}) it follows that

\begin{equation}
\cos \theta _2=\beta ^{-1}\left[ 1-\frac \omega \varepsilon (1-\beta \cos
\theta _1)+\frac \omega {mc^2\gamma }(1-\cos \theta )\right] .  \label{ct2}
\end{equation}
We are interesting in case $\gamma \gg 1$ and in all formulas that followed
we use the expansion $\beta =1-\frac 1{2\gamma ^2}$. In the spherical
coordinates $dV=2\pi r^2drd\cos \theta _2=$ $2\pi r^2\omega \left( 1-\cos
\theta _1\right) d\varepsilon dr/\varepsilon ^2$.

When $\varepsilon \gg \omega $, then $\theta _2\ll 1$ and thus 
\begin{equation}
\cos \theta _1\approx \cos \theta =\frac{r/a-\cos \psi }{\sqrt{\left(
r/a-\cos \psi \right) ^2+\sin ^2\psi }}.  \label{ct1}
\end{equation}

Now it is possible to rewrite (\ref{def}) as an integral over $r$ 
\[
L_\gamma \left( \omega ,\varepsilon ,\gamma ,\psi \right) = 
\]
\begin{equation}
\frac{N_\omega N_{e^{\pm }}}{8\pi a^2c}\frac \omega \varepsilon
\int\limits_0^{r_{\max }}\left( 1-\cos \theta _1\right) ^2\sigma _K\frac{dr}{%
\left( r/a-\cos \psi \right) ^2+\sin ^2\psi },  \label{def1}
\end{equation}
where $\,N_\omega $ is the number of photons going from the companion per
second and $N_{e^{\pm }}$ is the number of particles leaving the pulsar per
second, $n_\omega =N_\omega \left/ 4\pi cl^2\right. $, $n_{e^{\pm
}}=N_{e^{\pm }}/4\pi cr^2$, $r_{\max }$ is determined from the condition cos$%
\theta _2=1$. If Lorentz factor of a particle is constant then from (\ref
{ct2}), (\ref{ct1}) follows:

\begin{equation}
r_{\max }=a\left( \cos \psi +\frac{\sin \psi }2\left( z-\frac 1z\right)
\right) ,  \label{rmax}
\end{equation}
where

\[
z=\sqrt{\mu \lambda -1},\qquad \mu =\frac{4\omega \gamma ^2}\varepsilon
,\qquad \lambda =1-\frac \varepsilon {mc^2\gamma }. 
\]

In the case $K\ll 1$ in Thomson limit the Lorentz factor of the particles is
practically constant in the effective scattering volume. In the case $\omega
\gamma /mc^2>1$ this statement is also correct. In this case if the mean
free path of a particle is much bigger then the binary separation: $%
l_{free}=\sigma _{K_{full}}L_{*}/4\pi \omega a^2c\gg a$ ($\sigma _{K_{full}}$
is a full Klein-Nishina cross-section) then the influence of the field of
soft photons on the spectrum and flux of the relativistic particles may be
neglected. In the highly relativistic case ($\omega \gamma /mc^2\gg 1$) $%
\sigma _{K_{full}}\sim \sigma _Tmc^2/\omega \gamma $ and we note that from $%
K\ll 1$ it follows that $l_{free}\gg a$.

After a substitution of the expression (\ref{sigm}) in (\ref{def1}) and the
following integration under the assumption that $\gamma $ is constant we
obtain :

\begin{equation}
L_\gamma \left( \omega ,\varepsilon ,\gamma ,\psi \right) =\tau \frac{%
N_{e^{\pm }}N_\omega }{\gamma ^2}\frac \varepsilon \omega F\left( \omega
,\varepsilon ,\gamma ,\psi \right) ,  \label{Lg}
\end{equation}
where the constant $\tau =\frac{3\sigma _T}{128\pi ^2ca}$ and in the range 
\begin{equation}
\omega \ll \varepsilon \leq \varepsilon _{\max }=2\gamma ^2\omega \frac{%
1+\cos \psi }{1+\frac{2\omega \gamma }{mc^2}\left( 1+\cos \psi \right) }
\label{omax}
\end{equation}
the function $F\left( \omega ,\varepsilon ,\gamma ,\psi \right) $ is
determined as follows:

\begin{eqnarray}
F &=&\frac 1{\sin \psi }\left[ \left( \lambda +\frac 1\lambda \right) \left(
2\arctan z-\psi \right) \right.  \label{Ff} \\
&&-\left. \frac 8{3\mu ^2\lambda ^2}\left( 2z^3-3z^2\tan \frac \psi 2+\tan
^3\frac \psi 2\right) \right] ,  \nonumber
\end{eqnarray}
and is equal to zero when $\varepsilon >\varepsilon _{\max }$. When $\omega
\ll \varepsilon \ll \varepsilon _{\max }$ the equation (\ref{Ff})
simplifies: 
\[
F=2\frac{\pi -\psi }{\sin \psi }. 
\]

In the Thomson limit $\varepsilon _{\max }=2\gamma ^2\omega (1+\cos \psi )$
and hence $1-\lambda =\frac \varepsilon {mc^2\gamma }<2\frac{\omega \gamma }{%
mc^2}(1+\cos \psi )\ll 1$. Under these conditions equation (\ref{Ff}) becomes

\begin{eqnarray}
F_T &=&\frac 1{\sin \psi }\left[ 4\arctan z-2\psi -\right.  \label{ft} \\
&&\ \left. \frac 8{3\mu ^2}\left( 2z^3-3z^2\tan \frac \psi 2+\tan ^3\frac
\psi 2\right) \right] ,  \nonumber
\end{eqnarray}
\[
z=\sqrt{\frac{4\omega \gamma ^2}\varepsilon -1}. 
\]
If we integrate equation (\ref{Lg}) with $F=F_T$ over the energy of the
scattered photons $\varepsilon $ we receive the $\psi $-dependence of the
total outgoing energy in according with the equation (\ref{de1}). The
relationship between functions $\Phi \left( \psi \right) $ and $F_T$ is: 
\[
\Phi \left( \psi \right) =\frac 38\int\limits_0^{\varepsilon _{\max }}\frac
\varepsilon {\omega \gamma ^2}F_T\frac{d\varepsilon }{\omega \gamma ^2} 
\]

Under the real conditions the spectrum of the optical photons is not
monochromatic. If the companion's radiation is similar with a radiation from
the black body with temperature $T$, then the number of soft photons in the
energy range from $\omega $ to $\omega +d\omega $ is : 
\[
dN_\omega =\frac{15L_{*}}{\pi ^4T^4}\frac{\omega ^2d\omega }{e^{\omega /T}-1}%
. 
\]
Replacing $N_\omega $ in formula (\ref{Lg}) with $dN_\omega $and integrating
over $\omega $ we receive: 
\begin{equation}
L_\gamma \left( \varepsilon ,\gamma ,\psi \right) =\frac{15\tau }{\pi ^4}%
\frac{N_{e^{\pm }}L_{*}}{\gamma ^2}\frac{\varepsilon F_{bb}\left(
\varepsilon ,\gamma ,\psi \right) }{T^2},  \label{lbb}
\end{equation}
\begin{equation}
F_{bb}\left( \varepsilon ,\gamma ,\psi \right) =\frac
1{T^2}\int\limits_0^\infty F\left( \omega ,\varepsilon ,\gamma ,\psi \right) 
\frac{\omega d\omega }{\left( e^{\omega /T}-1\right) }.  \label{fbb}
\end{equation}

Figure 3 shows the dependence of $F_{bb}$ on $\varepsilon $ for different
values of $\psi $ and $\gamma $. It can be seen that in the energy range
where the difference of the cross-section from the Thomson one is
significant the form of the spectrum changes considerably. Figure 3 also
illustrates the dependence of the maximum energy of the scattered photons $%
\varepsilon _{\max }$ on $\gamma $ ( according to (\ref{omax}) within the
Thomson limit $\varepsilon _{\max }=2\omega \gamma ^2(1+\cos \psi )$ and in
the opposite case $\frac{\omega \gamma }{mc^2}(1+\cos \psi )\gg 1$ the
maximum energy $\varepsilon _{\max }=2mc^2\gamma $ ). The $\psi -$dependence
of the energy gone with the scattered photons in a unit of time in a unit
solid angle can be seen in Figure 4. The energy is normalized to $C=\frac{%
15\tau }{\pi ^4}N_{e^{\pm }}L_{*}\gamma ^2$. 

During the orbital motion $\psi $ changes periodically. In Figure 5 the
dependence of $F_{bb}$ on $\varepsilon $ for different positions of the
companions in the circular orbit and for different values of $\gamma $ is
shown. In this figure $i$ is an inclination angle and $\varphi $ is a true
anomaly. The relation between $\psi $ and $\varphi $ is: $\cos \psi =\sin
i\cos \varphi $. Figure 6 shows the orbital angle $\varphi $ dependence of
the luminosity $L_\gamma (\varphi )$ going in the direction of the observer
for the different values of inclination angles and Lorentz factors.

Formulas (\ref{rmax}) - (\ref{fbb}) were received under the assumptions that
Lorentz factor is constant. If we consider in the Thomson limit the
dependence of $\gamma $on $r$ according to formula (\ref{gam}) then from
formula (\ref{def1}) the luminosity of the scattered radiation is: 
\begin{equation}
L_\gamma \left( \omega ,\varepsilon ,\gamma _0,\psi \right) =\tau N_{e^{\pm
}}N_\omega \frac \varepsilon {\omega \gamma _0^2}F_\gamma \left( \omega
,\varepsilon ,\gamma _0,\psi \right) ,
\end{equation}
where 
\begin{equation}
F_\gamma \left( \omega ,\varepsilon ,\gamma _0,\psi \right) =
\end{equation}
\[
\int\limits_0^{r_{\max }}\frac{\gamma _0^2}{\gamma ^2}\frac{1+\left[ \gamma
^{-2}\left( 1-\cos \theta _2\right) ^{-1}-1\right] ^2}{\left( r/a-\cos \psi
\right) ^2+\sin ^2\psi }\frac{dr}a. 
\]
The dependences of $r_{\max }$ and $\cos \theta _2$ on $r$ are determined by
the formulas (\ref{ct2}),(\ref{ct1}).

Figure 7 represents spectra of the scattered radiation obtained in the
Thomson limit for different values of the parameter $K=\gamma _0/\gamma _{*}$
with $\gamma _0=10^4$. Spectra are calculated taking into account the
dependence of $\gamma $ on $r$. Curve $K=0$ corresponds to the case of
constant Lorentz factor, which is described by formula (\ref{ft}). It can be
seen from this figure that the bigger is the value of parameter $K$ at a
given angle, the softer is the spectrum of the hard radiation. This figure
also illustrates that with the growth of parameter $K$ the total energy of
the outgoing radiation increases tending to $L_w$as $K\rightarrow \infty $.

\section{Discussion}

We considered the scattering of the soft photons emitted by the optical star
by the relativistic electrons and positrons from the pulsar wind. We showed
that the intensity of the hard radiation came about from such a scattering
is proportional to the luminosities of the pulsar and the optical star. But
it is worth to mention that the luminosity of optical star itself depends on
the luminosity of the pulsar as it absorbs and reradiate part of the energy
of the pulsar wind. If we denote by $L_{st}$the intrinsic ( un-illuminated)
luminosity of the star and by $L_{ind}=fL_pR_{*}^2/4a^2$the energy of the
pulsar reradiated by the optical star ( $f$is the fraction of the energy
flux from the pulsar incident on the surface of the optical star that is
converted into soft emission) then the total luminosity of the star is $%
L_{tot}=L_{st}+L_{ind}$. PSR 1957-20 is an example of the system where the
illumination of the optical star by a pulsar affects the luminosity of the
optical star. A detailed study of the optical light curve of PSR 1957+20
done by Callanan et al. (1995) shows that the secondary is close to fill its
Roche lobe with 7\% - 20\% of the incident flux converted to optical
emission. Thus the described mechanism of the generation of the hard
radiation is important not only in case of a highly luminous optical star in
binary but also in case of a close binary with bright pulsar and an optical
star with a big radius.

\subsection{Comparison with the observations.}


The only known binary system which contains radio pulsar and emits non
pulsed X-ray radiation is PSR B1259-63 system. Since its discovery, the PSR
B1259-63 system has been observed several times at X-ray energies. ROSAT
observed the PSR B1259-63 system near to apastron in 1991-1992 (Cominsky et
al. 1994, Greiner et al. 1995). In January 1994 the X-ray emission from the
system PSR B1259-63 during the periastron passage was observed by telescopes
ASCA and OSSE (Grove et al. 1995, Kaspi et al. 1995). ASCA also observed the
post-periastron emission from this system on February 28, 1994.

The PSR B1259-63 system contains the radio pulsar with the spin period $%
P=47.76$ms, rotating around the massive Be star SS 2883. According to
Manchester et al., 1995 the orbital eccentricity is $e=0.87$, the projected
semimajor axis is $a_{mj}\sin i=3.9\times 10^{13}$cm and the longitude of
periastron is $\omega =139^{\circ }$. Spin-down luminosity of the pulsar is $%
L_p\simeq 9\times 10^{35}$erg/s. The mass function $f\left( M_p\right) =%
\frac{\left( M_c\sin i\right) ^3}{\left( M_p+M_c\right) ^2}=1.53M_{\odot }$,
the luminosity of the Be star is $L_{*}=2.2\times 10^{38}$erg/s (Johnston et
al. 1992, 1994). The distance between companions at periastron is $%
a=a_{mj}(1-e)\sim 10^{13}$cm.

The review of the different discussed models of the origin of the non pulsed
X-ray spectrum from the system can be found in the paper of Tavani\&Arons
1997. We tried to apply the results of our paper to this system. Assuming
the power law distribution of the electrons and positrons in the pulsar wind 
$\frac{dN_{e^{\pm }}}{d\gamma }=C\gamma ^{-s}L_w/mc^2$, $C=\left[
(s-2)\left/ \left( \gamma _{\min }^{2-s}-\gamma _{\max }^{2-s}\right)
\right. \right] $ and rewriting formulas (\ref{lbb}), (\ref{fbb}) taking
into the account the dependence $N_{e^{\pm }}(\gamma )$we receive

\begin{equation}
L_\gamma \left( \varepsilon ,\psi \right) =\frac{15\tau }{\pi ^4}\frac{%
CL_wL_{*}}{mc^2}\frac{\varepsilon F_{av}\left( \varepsilon ,\psi \right) }{%
T^2},  \label{fav}
\end{equation}
\begin{equation}
F_{av}\left( \varepsilon ,\psi \right) =\int\limits_{\gamma _{\min
}}^{\gamma _{\max }}F_{bb}\left( \varepsilon ,\gamma ,\psi \right) \gamma
^{-s-2}d\gamma .
\end{equation}
The result of our model for $10<\gamma <500$, $s=2.4$ and the measured
spectrum are shown in Figure 8. Spectrum measured by OSSE (black dots) was
taken from the paper of Grove et al. 1995. The straight lines shows the
extrapolation of ASCA data (Kaspi et al. 1995).As it can be seen from the
Figure 8 while the form of the spectrum is quite close to the observed one
the intensity of the radiation resulted from our model is less then the
observed one by a factor of 30.

This discrepancy is due to the fact that we derive the formula (\ref{fav})
under the assumption of the undisturbed free flow of the pulsar wind which
is not correct in the system PSR B1259-63 where the strong mass outflow from
the Be star presents. Writing the mass rate of the polar wind of the Be star
as $\stackrel{\cdot }{M}=(10^{-8}M_{\odot }/yr)\stackrel{\cdot }{M}_{-8}$and
the wind velocity as $v=10^8v_8$cm/s we find the ratio of the impulses of
the two winds $L_p/cv\stackrel{\cdot }{M}=0.45/v_8\stackrel{\cdot }{M}_{-8}.$
Thus for the Be star polar wind typical parameters the centrally located
shock between the pulsar and the star due to the interaction between the two
winds seems to appear and thus the free flow of the pulsar wind will be
disturbed. The big differences between the values of the velocities of the
particles from the different sides of the tangential discontinuity will lead
to the growth of the instabilities and the two winds will be macroscopically
mixed between the shocks. The results of numerical calculations of
Igumenshchev (1997) verified such a picture. Then the heavy non relativistic
wind slows down the volumes filled by the relativistic electrons and
positrons and they acquire essentially non relativistic hydrodynamic drift
velocity $v_{d}$ along the shock while the energy of electrons and
positrons does not changes significantly. With the decrease of the
hydrodynamic velocity of the relativistic plasma the time which it spends
near the optical star increases in $c/v_d$ times. The idea of small drift
velocity as applied to the system $LSI$ $61^{\circ }303$ was discussed by
Maraschi\&Treves (1981) The effective transformation parameter $K_{eff}\sim
\frac c{v_d}K$ thus can be large enough to overcome the discrepancy between
the simple theory and observations.

Under the assumption of the power law energy distribution of the
relativistic electrons and positrons after the shock $\frac{dN_{e^{\pm }}}{%
d\gamma }=A\gamma ^{-s}$ and their isotropic velocity distribution we can
estimate the intensity of the X-ray radiation from the system. Such a
distribution can be either the result of the particle acceleration on the
shock or in the case of the intrinsic power law distribution of the
electrons and positrons in the pulsar wind and thin collisionless shock,
passing which the relativistic particles do not change their energy
distribution.

Lets consider an element of volume $dV$ filled with the relativistic plasma
locating in the shock on a distance $R$ from the Be star. In the case of the
relativistic electrons and positrons isotropic distribution the number of
particles produces the photons with energy $\varepsilon $ moving in the
direction of the observer in a unit solid angle is $\frac{dN_{e^{\pm }}}{%
d\gamma }\frac{d\Omega }{4\pi }d\gamma =0.5\frac{dN_{e^{\pm }}}{d\gamma }%
d\cos \theta _2d\gamma =\frac 12\frac{dN_{e^{\pm }}}{d\gamma }\frac \omega
{\varepsilon ^2}\left( 1-\cos \theta \right) d\varepsilon d\gamma $ (see
section 2.2). Then according to (\ref{def}, \ref{sigm}, \ref{enscat}) the
number of photons coming to the observer in the unit of time per unit square
per MeV is

\[
dN=\frac{3\sigma _T}{32\pi }\frac{An_\omega c}{\omega D^2}\times 
\]
\begin{equation}
\int\limits_{\sqrt{\varepsilon /2\omega (1-\cos \theta )}}^\infty \gamma
^{-s-2}\left[ 1+\left( 1-\frac \varepsilon {\omega \gamma ^2(1-\cos \theta
)}\right) ^2\right] d\gamma dV  \label{dn}
\end{equation}
where $\theta $ is a photon scattering angle. For the simplicity we take the
photon density along the shock constant and equal to the one at a distance $%
R=10^{13}cm$. The factor $AV$can be estimated from the energy conservation
law by equating the energy enters the shock per second $L_p\Omega /4\pi $and
the energy leaving the shock per second $AVmc^2v_d\gamma _{min}^{2-s}/a(s-2)$%
, where $\Omega $ is the solid angle under which the shock wave is seen from
the pulsar. Thus we have $AV\sim L_p\Omega a(s-2)/4\pi mc^2v_d\gamma
_{min}^{2-s}$. Then integrating (\ref{dn}) over the volume of the shock we
have for the $N$ 
\begin{equation}
N\sim 2\tau \frac{a^2}{D^2R^2}\frac{L_{*}L_p}{mcv_d}\frac \Omega \pi \frac{%
(s-2)}{\gamma _{min}^{2-s}}\frac{(2\omega )^{(s-3)/2}}{(1+s)}\varepsilon
^{-(1+s)/2}.
\end{equation}
For $\gamma _{min}=10,s=2.4,D=2$kpc$,v_d=10^8$cm/s we have 
\begin{equation}
N\sim 5\times 10^{-3}(\frac \varepsilon {100keV})^{-1.7}ph/s/cm^2/MeV
\end{equation}
while the observable value of the radiation at a 100 keV is $2.8\times
10^{-3}ph/s/cm^2/MeV$. Thus if the small drift velosity is taken into
account it is possible to explain the observed spectral shape and intensity
of the X-ray radiation. It is also worth to mention that for the particles
with big Lorentz factor $\gamma >\gamma _{*}v_d/c$ the assumption of
constant Lorentz factor will be not valid due to the inverse Compton losses
and the break in the photon spectrum will appear. The break in the photon
spectrum at $\varepsilon =\omega \gamma _{*}^2v_d^2/c^2$ will appear. The
index of the photon spectrum after the break will be bigger then the
original one at 1/2. It can be also seen from (\ref{dn}) that the focus
effect will take place - the intensity of the radiation is proportional to
the $(1-\cos \theta )^{(1+s)/2}$and thus the major part of the radiation
will be emitted toward the direction of the optical star.

\vspace{1cm} {\bf ACKNOWLEDGMENTS} \vspace{0.5 cm}

We are grateful to V.M.Kaspi and M.Tavani for valuable comments and to
V.S.Beskin and Ya.N.Istomin for helpful discussions. This work is supported
in part by the RFBR grant 97-02-16975 and the Cariplo Foundation for
Scientific Research.

\pagebreak
{\bf Figure Captions}

1)The geometry under consideration. $P$ is a pulsar, $S$ is an optical star, 
$I$ is a point of interaction of the electrons and photons, $O$ is an
observer. Note that these points are not in one plane in general case.

2)The dependence of the beaming function $\Phi $ on an angle $\psi .$

3)The dependence of $F_{bb}$on $\varepsilon $ for different values of $\psi $
for $\gamma =10^4$and $\gamma =10^6$.

4)The $\psi $-dependence of the energy gone with the scattered photons in a
unit of time in a unit solid angle.The energy is normlized to $C=\frac{%
15\tau }{\pi ^4}N_{e^{\pm }}L_{*}\gamma ^2.$

5)The dependence of $F_{bb}$on $\varepsilon $ for different positions of the
companions in the circular orbit for $\gamma =10^4$and $\gamma =10^6$.

6)The orbital angle $\varphi $ dependence of the total energy gone with the
scattered quanta in the unit of time in the unit solid angle for the
different values of inclination angles a)$\gamma =10^4$, b)$\gamma =10^6.$

7)Spectra of the scattered radiation obtained for different values of the
parameter $K={\gamma _0}/{\gamma }_{*}$. Spectra are calculated taking into
account the dependence of $\gamma $ on $r$.

8)We apply our model to the system PSR B1259-63. Theoretical result (solid
line), OSSE spectrum of emission from 1994 January 3-23(black dots) and
schematic extrapolations to the power-law spectra of {\it ASCA} observations
from 1993 December 28(dashed line), 1994 January 10 (dotted line), and 1994
January 26 (dashed-dotted line)are shown. {\it ASCA} extrapolations obtained
from the analysis of Kaspi et al. 1995, OSSE results are taken from the
paper of Grove et al. 1995.

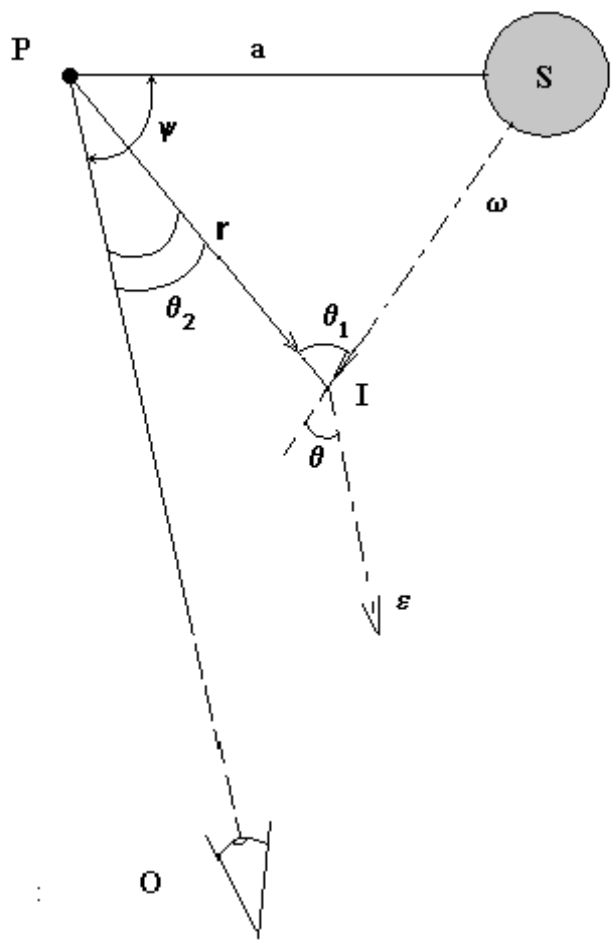
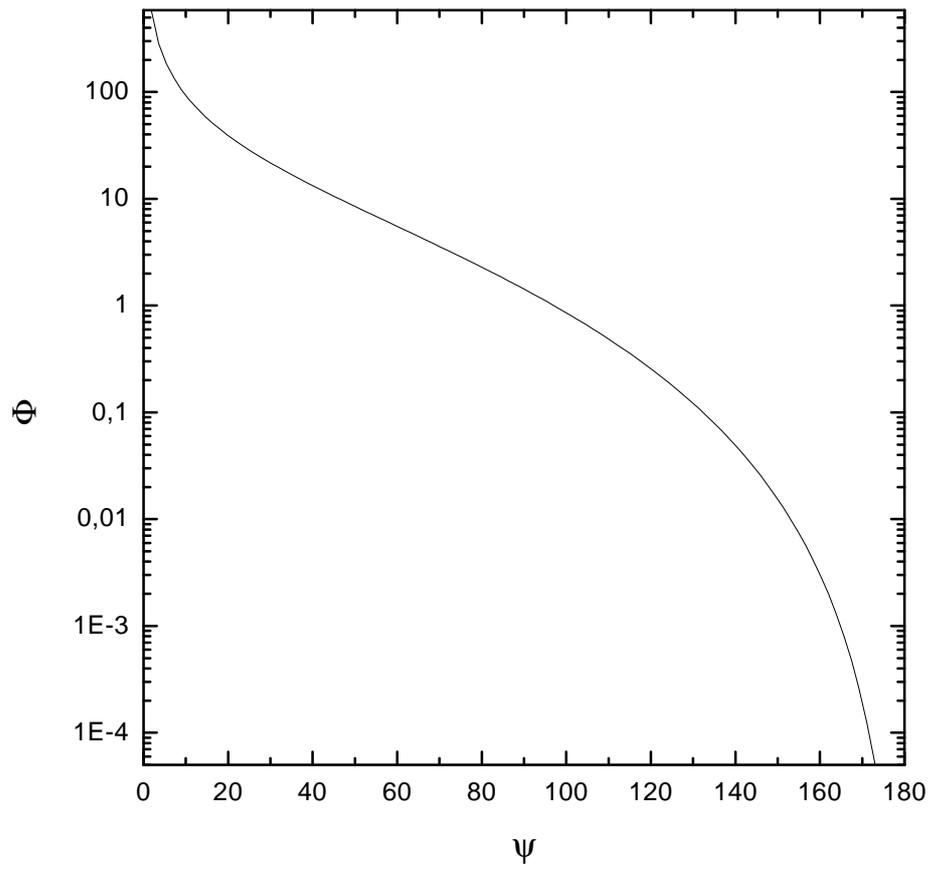
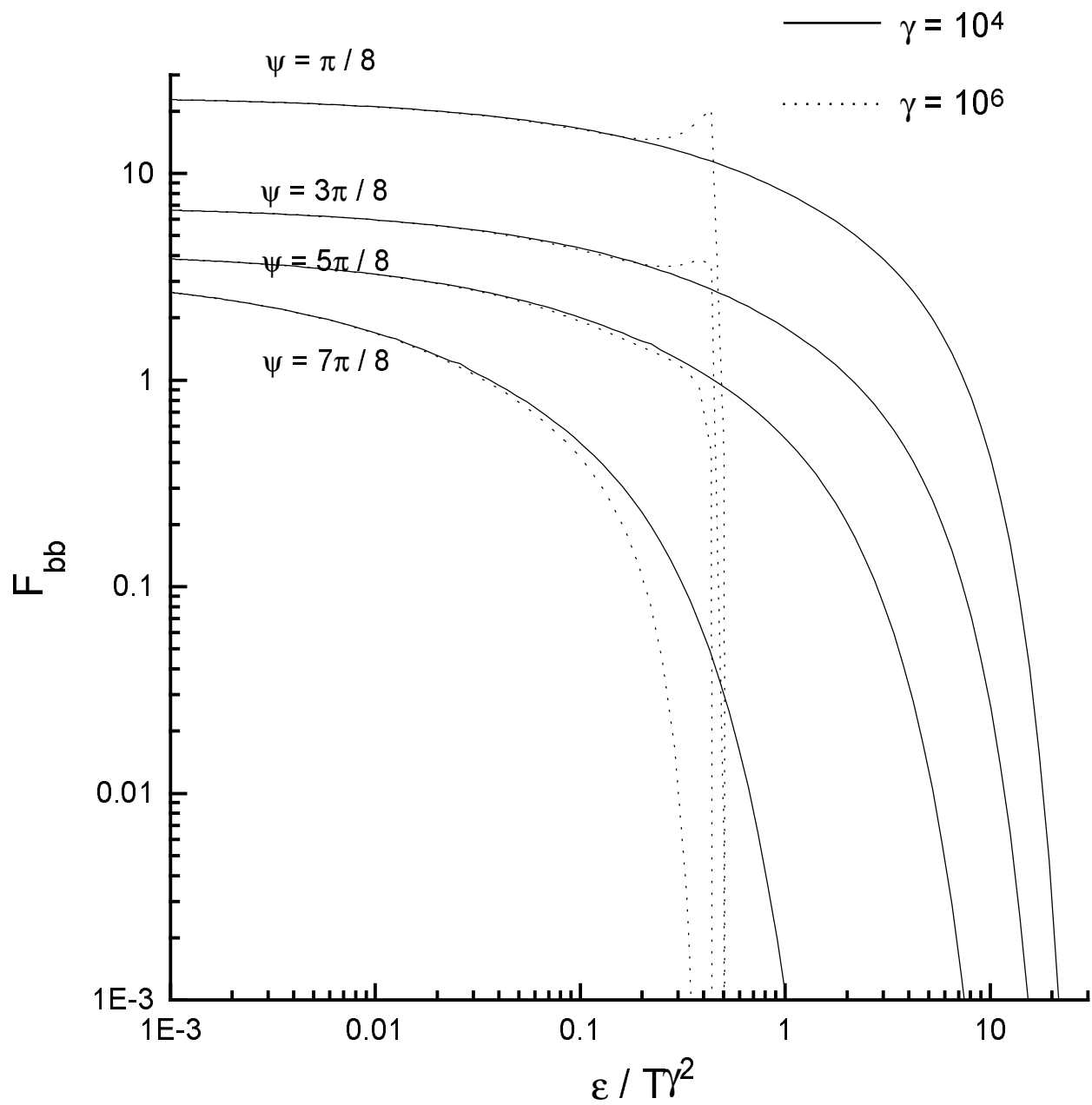
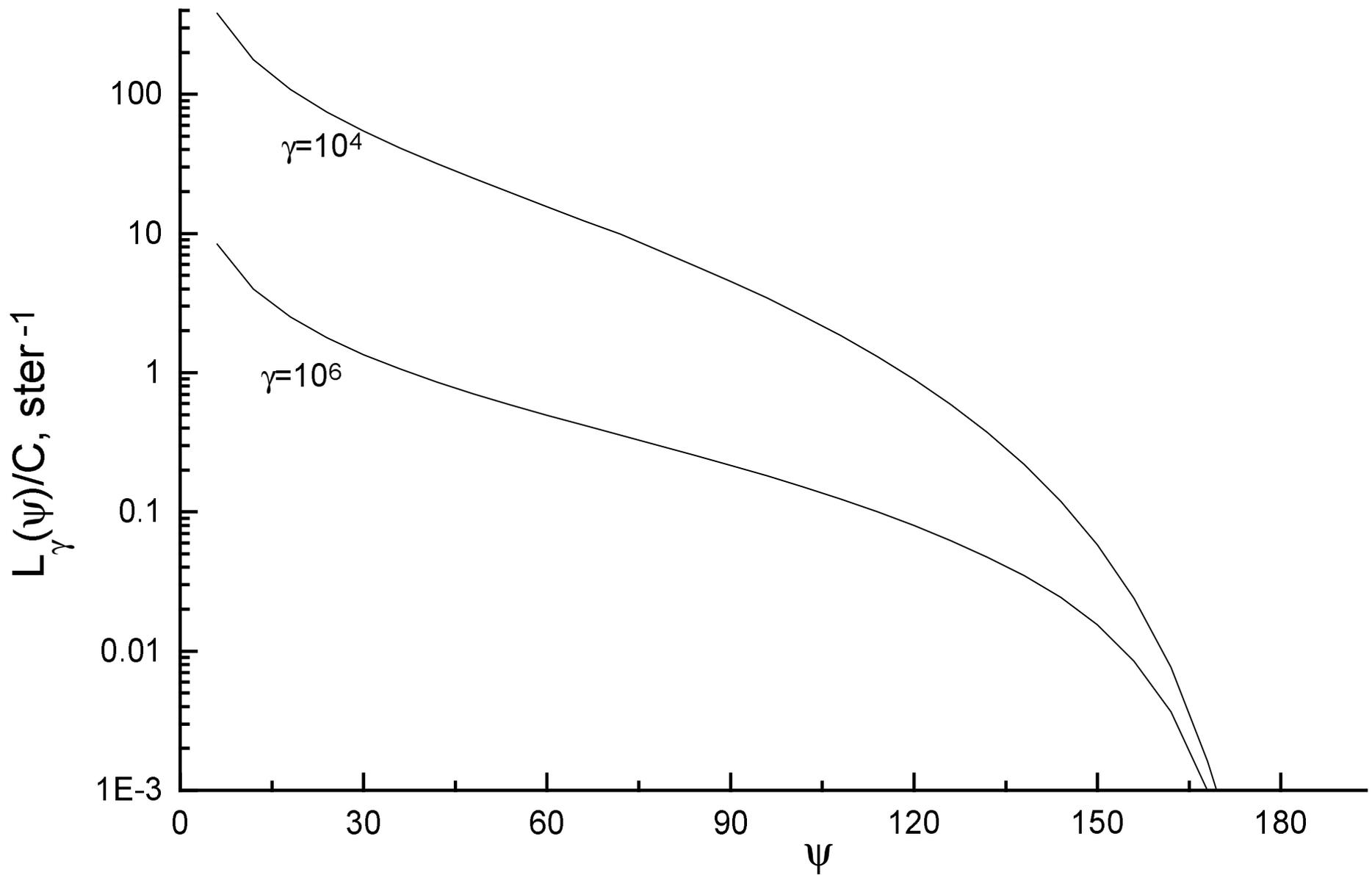
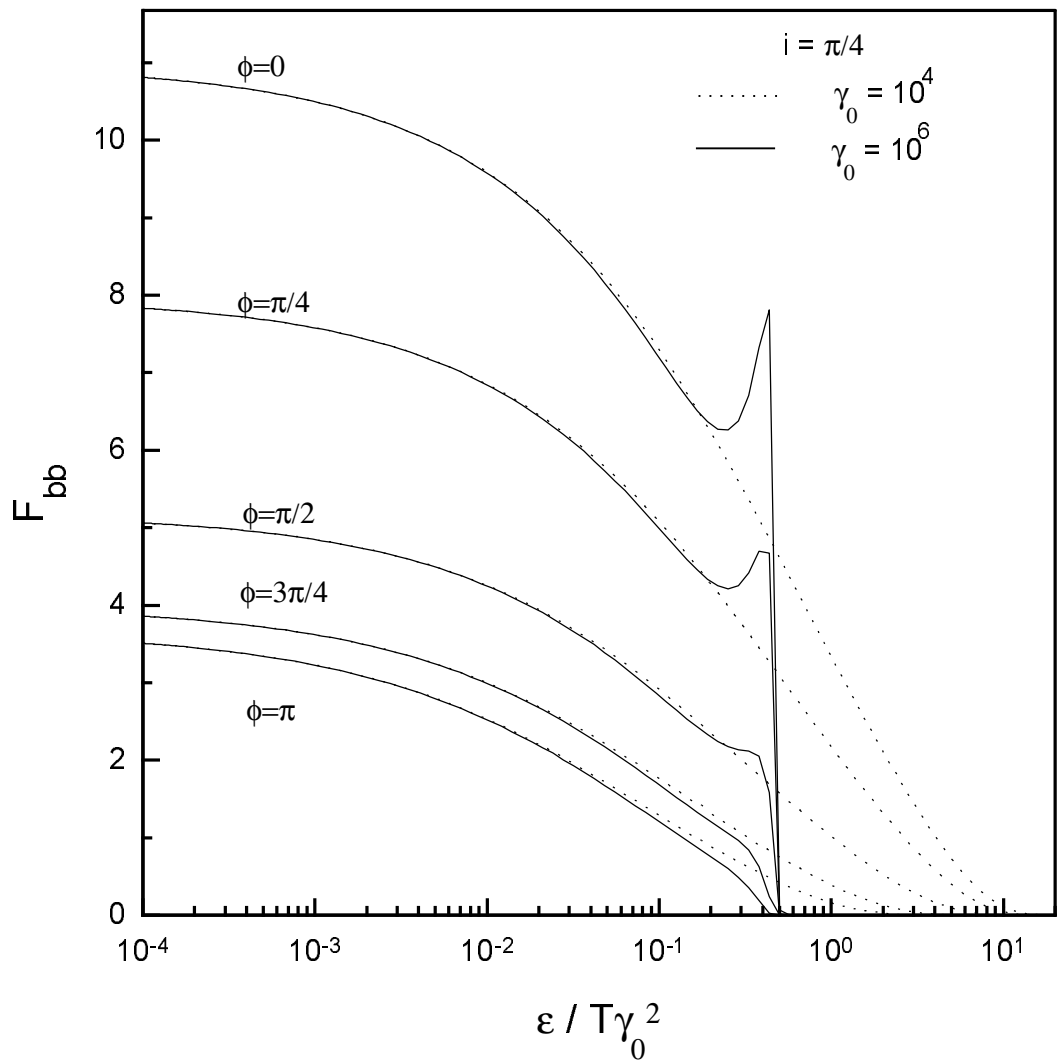
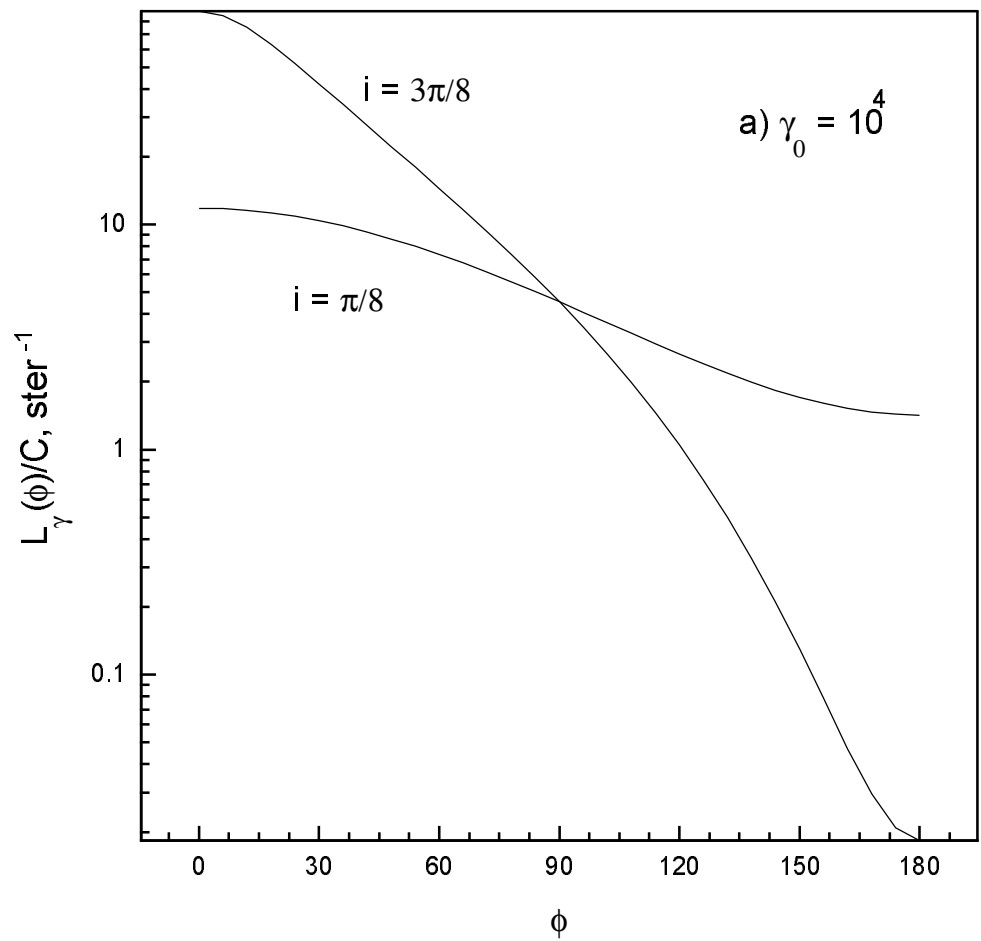
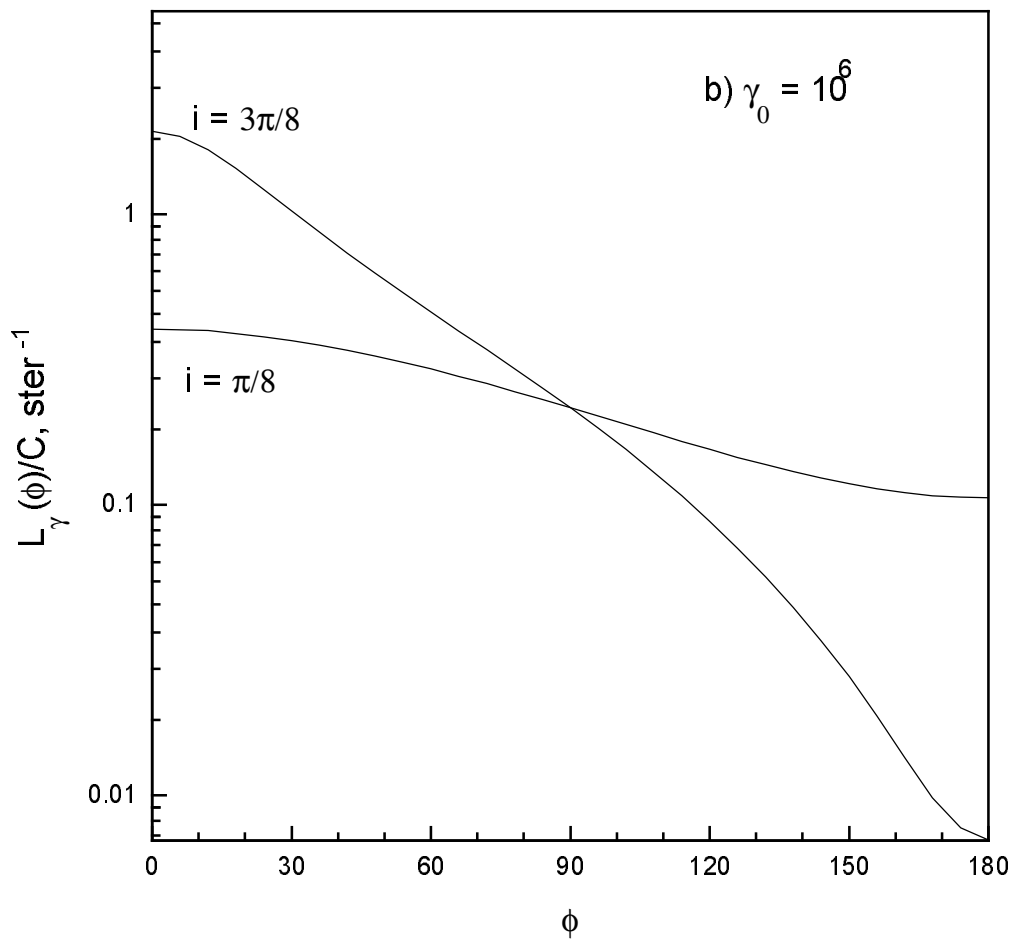
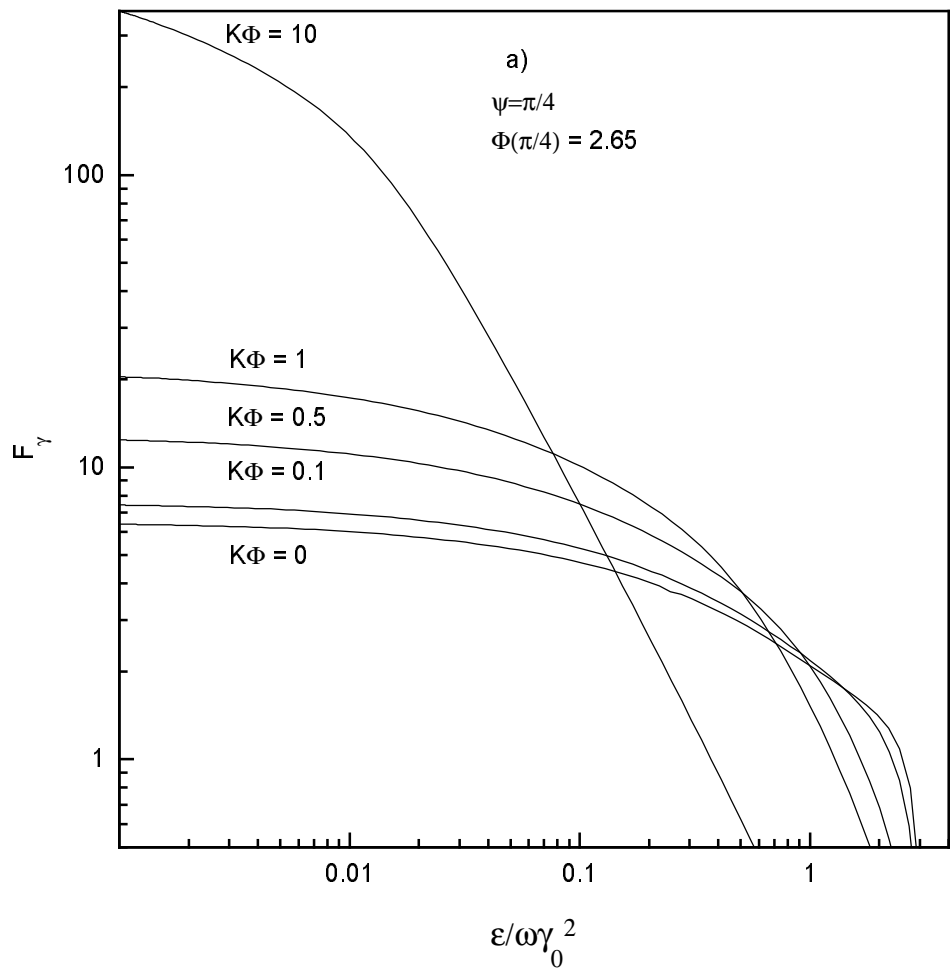
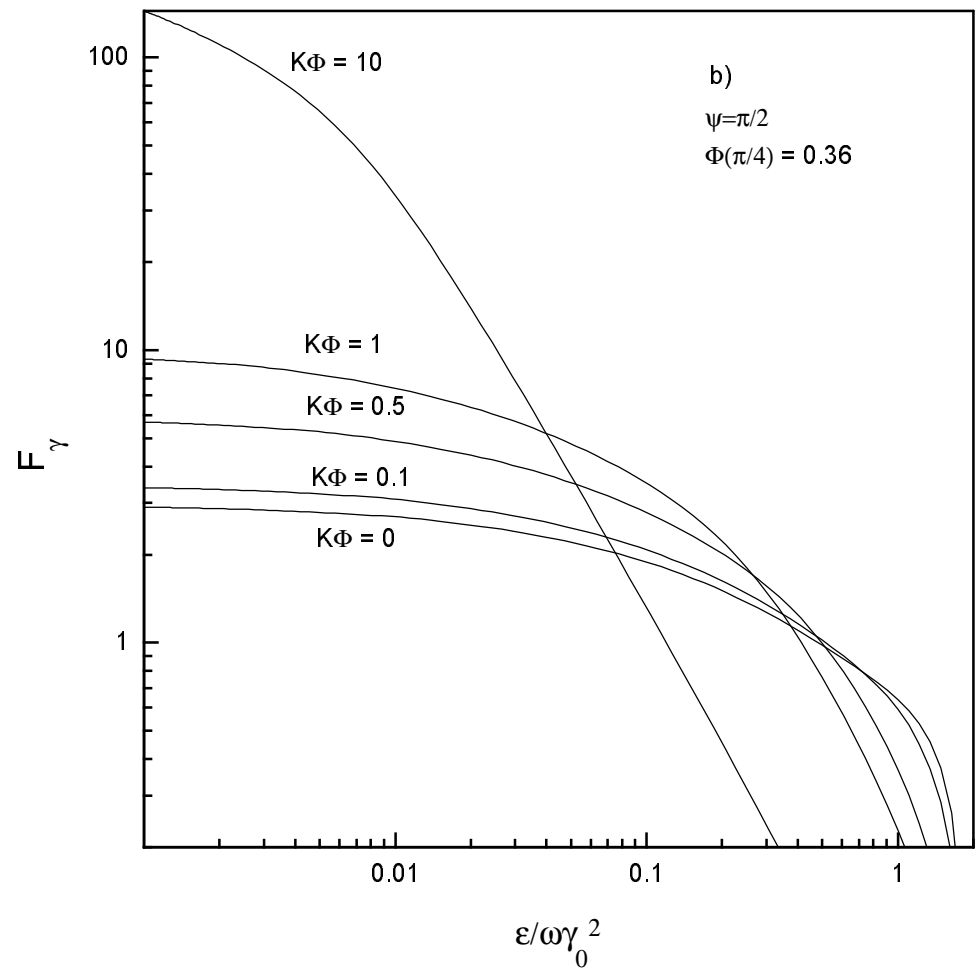
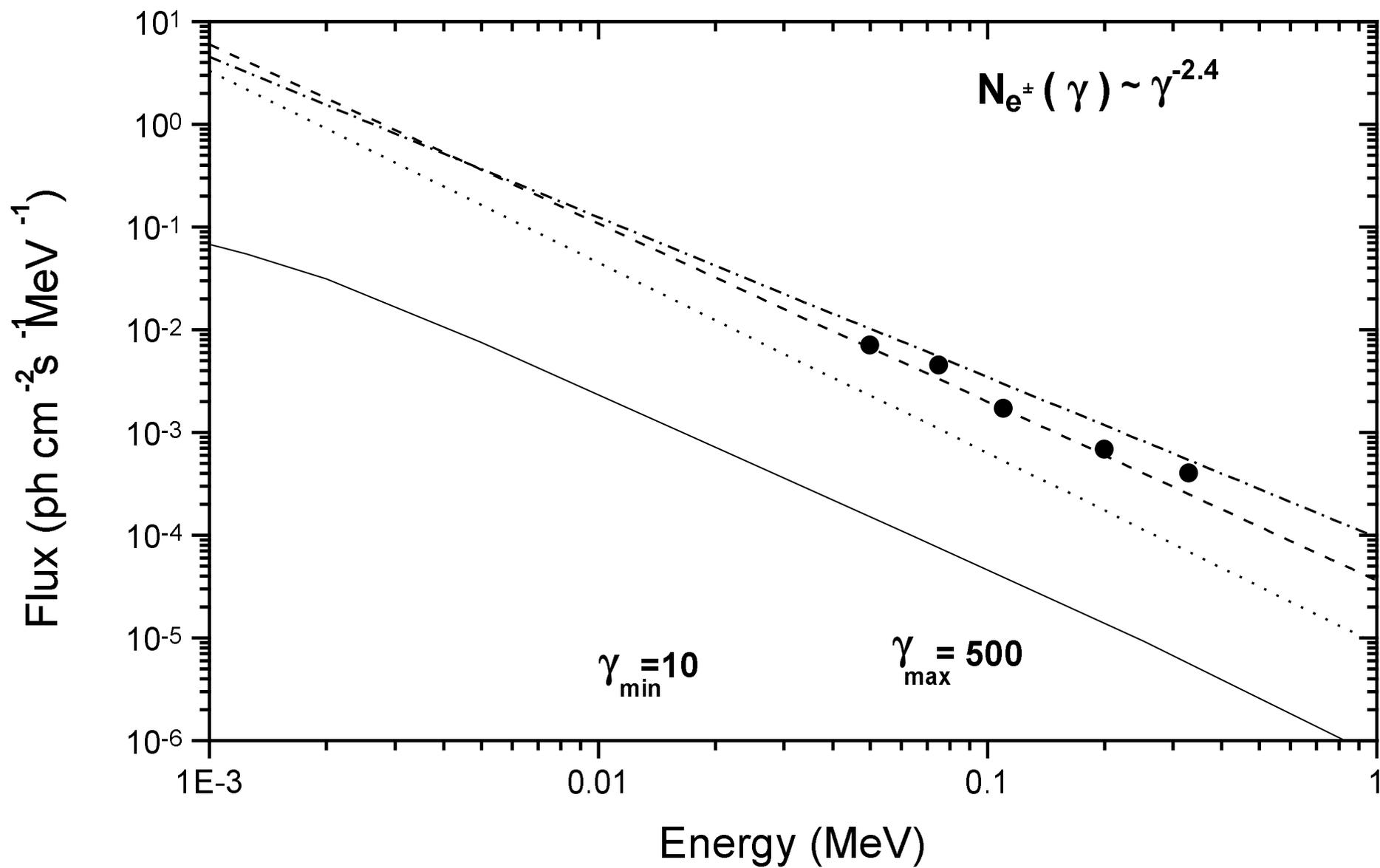
\end{document}